\newdimen\rotdimen
\def\vspec#1{\special{ps:#1}}
\def\rotstart#1{\vspec{gsave currentpoint currentpoint translate
   #1 neg exch neg exch translate}}
\def\rotfinish{\vspec{currentpoint grestore moveto}}
\def\rotr#1{\rotdimen=\ht#1\advance\rotdimen by\dp#1%
   \hbox to\rotdimen{\hskip\ht#1\vbox to\wd#1{\rotstart{90 rotate}%
   \box#1\vss}\hss}\rotfinish}
\def\rotl#1{\rotdimen=\ht#1\advance\rotdimen by\dp#1%
   \hbox to\rotdimen{\vbox to\wd#1{\vskip\wd#1\rotstart{270 rotate}%
   \box#1\vss}\hss}\rotfinish}%
\def\rotu#1{\rotdimen=\ht#1\advance\rotdimen by\dp#1%
   \hbox to\wd#1{\hskip\wd#1\vbox to\rotdimen{\vskip\rotdimen
   \rotstart{-1 dup scale}\box#1\vss}\hss}\rotfinish}%
\def\rotf#1{\hbox to\wd#1{\hskip\wd#1\rotstart{-1 1 scale}%
   \box#1\hss}\rotfinish}%
\newbox\rotbox

\documentstyle[twoside,fleqn,espcrc2,epsf]{article}

\newcommand{\AmS}{{\protect\the\textfont2
  A\kern-.1667em\lower.5ex\hbox{M}\kern-.125emS}}

\title{Chiral Gauge Theories and Fermion--Higgs Systems}

\author{Donald N. Petcher\address{Department of Physics, Washington University,
        Saint Louis, Missouri 63130 USA}%
        }

\newcommand{\MFLG}{M.\ F.\ L.\ Golterman}
\newcommand{\DNP}{D.\ N.\ Petcher}
\newcommand{\JS}{J.\ Smit}

\newcommand{\be}{\begin{equation}}
\newcommand{\ee}{\end{equation}}
\newcommand{\bea}{\begin{eqnarray}}
\newcommand{\eea}{\end{eqnarray}}
\newcommand{\ba}{\begin{array}}
\newcommand{\ea}{\end{array}}
\newcommand{\nn}{\nonumber}
\newcommand{\eq}[1]{eq.~(\ref{eq:#1})}
\newcommand{\half}{\mbox{\small $\frac{1}{2}$}}

\newcommand{\V}[1]{V_{#1}}
\newcommand{\Vdagger}[1]{{V^\dagger_{#1}}}
\newcommand{\g}[1]{g_{#1}}
\newcommand{\gdagger}[1]{{g^\dagger_{#1}}}
\newcommand{\h}[1]{h_{#1}}
\newcommand{\hdagger}[1]{{h^\dagger_{#1}}}
\newcommand{\D}{{D}}
\newcommand{\Dtilde}{{\tilde{\D}}}
\newcommand{\Dmu}{{\D_\mu}}
\newcommand{\Dtildemu}{{\Dtilde_\mu}}
\newcommand{\DWmu}{{\D^{(W_\mu)}_\mu}}
\newcommand{\DWtildemu}{{\Dtilde^{(W_\mu)}_\mu}}
\newcommand{\latd}{{\partial}}
\newcommand{\latdtilde}{{\tilde \latd}}
\newcommand{\latdmu}{{\latd_\mu}}
\newcommand{\latdtildemu}{{\latdtilde_\mu}}
\newcommand{\PL}{{P_L}}
\newcommand{\PR}{{P_R}}
\newcommand{\tr}{{\rm tr}}
\newcommand{\hc}{\hbox{\rm h.c.}}

\newcommand{\muhat}{{\hat\mu}}

\newcommand{\pslash}{{\hbox{$p$}\kern-0.4em\raise0.10ex\hbox{$/$}}}
\newcommand{\meff}{M}
\newcommand{\sigmahatmu}{{\hat\sigma}_\mu}
\newcommand{\pr}[2]{P^#1_#2}
\newcommand{\xtilde}{{\tilde x}}
\newcommand{\Umu}[1]{{U_{\mu #1}}}
\newcommand{\Udaggermu}[1]{{U^\dagger_{\mu #1}}}
\newcommand{\ULmu}[1]{{U^L_{\mu #1}}}
\newcommand{\UdaggerLmu}[1]{{U^{L\dagger}_{\mu #1}}}
\newcommand{\URmu}[1]{{U^R_{\mu #1}}}
\newcommand{\UdaggerRmu}[1]{{U^{R\dagger}_{\mu #1}}}
\newcommand{\URLmu}[1]{{U^{RL}_{\mu #1}}}
\newcommand{\ULRmu}[1]{{U^{LR}_{\mu #1}}}
\newcommand{\UdaggerRLmu}[1]{{U^{RL\dagger}_{\mu #1}}}
\newcommand{\UdaggerLRmu}[1]{{U^{LR\dagger}_{\mu #1}}}
\newcommand{\PhiRLmu}[1]{{\Phi^{RL}_{\mu #1}}}
\newcommand{\PhiLRmu}[1]{{\Phi^{LR}_{\mu #1}}}

\newcommand{\PhidaggerLRmu}[1]{{\Phi^{LR\dagger}_{\mu #1}}}

\newcommand{\Wmu}[1]{W_{\mu#1}}

\newcommand{\psibar}{{\overline{\psi}}}
\newcommand{\psiR}[1]{\psi_{R#1}}
\newcommand{\psiL}[1]{\psi_{L#1}}
\newcommand{\psiLi}[1]{\psi^i_{L#1}}
\newcommand{\psibarL}[1]{\psibar_{L#1}}
\newcommand{\psibarR}[1]{\psibar_{R#1}}
\newcommand{\psin}[1]{\psi^{(n)}_{#1}}
\newcommand{\psibarn}[1]{\psibar^{(n)}_{#1}}
\newcommand{\psiRn}[1]{\psi_{R#1}^{(n)}}
\newcommand{\psiLn}[1]{\psi_{L#1}^{(n)}}
\newcommand{\psibarLn}[1]{\psibar_{L#1}^{(n)}}
\newcommand{\psibarRn}[1]{\psibar_{R#1}^{(n)}}
\newcommand{\psic}[1]{\psi^{(c)}_{#1}}

\newcommand{\psidaggerL}[1]{\psi^\dagger_{L#1}}
\newcommand{\psitransposeL}[1]{\psi^T_{L#1}}
\newcommand{\psistarL}[1]{\psi^*_{L#1}}
\newcommand{\chibar}{{\overline{\chi}}}
\newcommand{\Psibar}{{\overline{\Psi}}}

\newcommand{\Phidagger}{{\Phi^\dagger}}

\newcommand{\Phimu}[1]{\Phi_{\mu#1}}
\newcommand{\SWY}{S_{\rm WY}}
\newcommand{\SpWY}{S^\prime_{\rm WY}}
\newcommand{\SRoma}{S_{\rm Roma}}
\newcommand{\Sgf}{S_{\rm g.f.}}
\newcommand{\Sghosts}{S_{\rm ghosts}}
\newcommand{\Sct}{S_{\rm c.t.}}
\newcommand{\Seff}{S_{\rm eff}}
\newcommand{\Sgauge}{S_{\rm gauge}}

\begin{document}

\begin{abstract}
The status of several proposals for defining a theory of chiral fermions on
the lattice is reviewed and some new estimates for the upper bound on
the Higgs mass are presented.
\end{abstract}

\maketitle

\section{Introduction}

The amount of work on lattice theory relating to chiral gauge theory has
dramatically increased this year with some forty papers in the subject,
represented by about twenty contributions to this conference.
In contrast, only a handful of papers have been written in the broader
field of fermion--Higgs systems without an eye toward chiral gauge
theory. Therefore, with one exception, I have decided to
limit my review to the topic of chiral gauge theories, and within that I
will primarily emphasize the conceptual developments rather than numerical
simulations.  The exception is some work in the
pure Higgs sector concerning the regularization dependence of the upper
bound for the Higgs mass\cite{Heller} to which I would like to draw your
attention.

Last March, a very pleasant workshop was held in Rome, on the topic of
non-perturbative approaches to chiral gauge theories. In that workshop,
virtually all proposals around at the time were presented, following
which some quite intense discussions took place in the proverbial
smoke-filled back room\footnote{Although due to federal
regulations the smoke was not allowed} in an attempt to understand the
problems of some of the proposals and the connections between them. Some
valuable insights were gained in these discussions and in my talk I will
attempt to portray the underlying theme that we arrived at during that
meeting. The topics discussed in the Rome workshop are the following:
\begin{itemize}
\item the failure of the Wilson--Yukawa approach
\item consequences for the Eichten--Preskill approach
\item the Rome approach and its relation to the above
\item the staggered fermion approach
\item mirror fermions
\item Zaragoza ``replica'' fermions
\item the topological method
\item contribution of doublers to the S--parameter
\item the Banks ``U(1) problem''
\end{itemize}
\noindent
Suggestions that were not represented at the meeting
are:
\begin{itemize}
\item the Bodwin-Kovacs proposal
\item reflection positive fermions
\end{itemize}
which I would also like to mention, and finally I would like to spend
some time to
discuss a proposal made after the conference which is perhaps the
most interesting proposal presently on the table:
\begin{itemize}
\item domain wall fermions
\end{itemize}
Aside from these, a rather pessimistic
contribution concerning the use of the
random lattice as a solution to the chiral fermion problem
was presented at this conference\cite{Kieu}, although I am not prepared to
discuss this work.

Naturally with the large amount of material represented above, I have been
forced to make some subjective choices about what to cover and what to omit.
So keeping in mind my attempt to focus on the conceptual development of the
field, I have decided to omit discussion of mirror fermions\cite{Montvay}
for the most part with a twofold justification. First, that theory is a
vector theory in principle and not an attempt at producing a chiral theory.
The chiral theory is modeled at low energy, with the chiral partners of
these fermions appearing close to the electroweak scale. Second, and perhaps
more to the point, the theory has already been well represented in plenary
talks of the last two lattice conferences\cite{Phasediagram} with little
{\it conceptual} development since then. I would like to comment though that
Montvay gave us a nice talk in Rome indicating that the theory of
mirror fermions is not in contradiction to any known experimental phenomena
to date, which should force us not to foreclose on the possibility that
mirror fermions actually do exist in nature, and the world is not chiral
after all. Work goes on in the numerical arena, and as this is the first
year that the effects of fermions on the upper bound of the Higgs mass have
been reported, I will mention some results obtained from the mirror fermion
method when I come to this topic.

I will also say little about the Banks problem other than to point
out what it is and how it can be resolved in various of the
proposals, and I will not have time to discuss the contribution of doubler
fermions to the S--parameter\cite{DugRand}. All other topics
mentioned above will find their way into my talk, although in several
cases I will have time only to present the barest essentials.

\section{HIGGS MASS UPPER BOUND}

Before a discussion of chiral fermions I would like to briefly describe some
progress on a topic that falls under my jurisdiction: recent results on
regularization dependence of the upper bound to the Higgs mass. In past
years, calculations of the upper bound to the Higgs mass have concentrated
primarily on a theory with standard action and hypercubic
lattice\cite{LueWei,Kuti}, with the only deviation being studies on an $F_4$
lattice\cite{Hellerearly}. Fixing $f_\pi$ at its experimental value of 246
GeV (within the $\sigma$-model context), a typical value for the upper bound
of the Higgs mass in these studies was 640 GeV. This year, for the first
time the horizons have been substantially broadened, with investigations of
theories with actions including higher derivative terms, using the $1/N$
approximation as well Monte Carlo calculations\cite{Heller}.

A typical graph of the results of a calculation of the
Higgs mass in units of $f_\pi$ versus the Higgs mass in cutoff units is
given in figure 1. These graphs represent calculations on an
$F_4$ lattice, a hypercubic lattice with standard action, a hypercubic
lattice with a Symanzik improved action, and a calculation with
Pauli--Villars regularization, as well as results from actions including
higher dimension operators. The
results from the $1/N$ expansion (extrapolated to $N=4$)
with the standard action and an action
including higher dimension operators are given in each case by the dotted
and solid lines respectively.
\begin{figure}[t]
\epsfxsize=\columnwidth
\epsffile{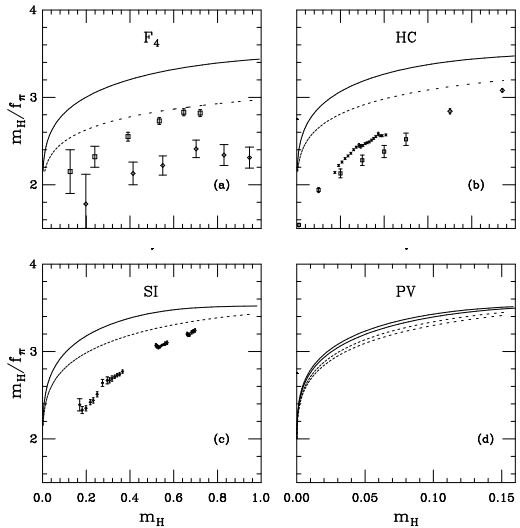}
\caption{Graphs containing numerical data and results from large $N$
calculations for
hypercubic (HC), $F_4$ and Symanzik improved lattice regularization schemes,
and with Pauli--Villars (PV) regularization, including results from actions
with dimension $6$ operators.}
\end{figure}
The main point I would like to make concerning this graph is that both
adding `irrelevant' terms to the action which represent cutoff effects, and
looking at a broader class of regularization schemes, allows the upper bound
on the Higgs mass to be relaxed.  With the usual
requirement that the cutoff effects are only a few per cent on pion
scattering, from this analysis a conservative estimate
for the upper bound is
about 750 GeV $\pm$ 50 GeV which is somewhat higher than the old
bound. In addition, the decay width is found to be about
290 GeV which is somewhat broader than the tree level value of 210 GeV,
indicating that in the region in question the Higgs
particle may be more strongly coupled than previously believed.  I will
come back to this topic
shortly when I consider the effects of fermions on these numbers.

\section{THE BANKS ``U(1) PROBLEM''}

Up to a little over a year ago, most models which were under investigation
as theories of chiral fermions possessed a rigid $U(1)$
symmetry corresponding to fermion number which commutes with the gauge
symmetry. Banks pointed out that for a realistic theory which hopes to
reproduce the standard model either directly or as a low energy
approximation to some other chiral theory obtained
from the lattice, this presented a problem from the point of view of
anomalies\cite{Banks}. For in the standard model, baryon number $B$ and lepton
number $L$ are each independently violated by instanton processes, and
only the difference $B-L$ is conserved, whereas the rigid $U(1)$
symmetry in the lattice theories would insure that both $B$ and $L$
would be conserved independently.  Although this problem had been
overlooked for the most part\footnote{It should be pointed
out that Eichten and Preskill did address this problem directly in their
approach.} in the face of other more severe problems, the comments by
Banks did force the rest of us to `come clean' and solutions to the
problem for the various models formed a part of our discussion in the
Rome workshop. There are several ways around the problem that are either
built in explicitly or realized dynamically by the various models which
I will point out as we go along.

\section{WILSON--YUKAWA, EICHTEN--PRESKILL AND ROMA}

There is very strong evidence by now that the Wilson--Yukawa approach fails
to result in a theory of lattice chiral fermions, and that unfortunately its
failure takes down the Eichten--Preskill approach with it. In this section I
review why this is so, and indicate how the findings are related to the
Rome approach and why the latter may escape the problems that arise in the
first two.

\subsection{The Wilson--Yukawa approach}

To understand the problems involved, it is sufficient to consider a
simple case of the Wilson--Yukawa model containing only a
single species of fermion, and in which only the left-handed
chiral symmetry is gauged, leaving the right--handed chiral symmetry rigid.
The model is defined by the fermion action\cite{Karstenearly,SmitZak,Swift}
\bea \SWY &=&
\sum_{x,\mu}\half\psibar_{x}\gamma_\mu(\Dmu+\Dtildemu)\psi_{x} \nn\\
&+&y\sum_x [\psibarL{x}\V{x}\psiR{x}+\psibarR{x}\Vdagger{x}\psiL{x}]
\label{eq:SSaction}\\
&-&\frac{w}{2}\sum_x [\psibarL{x}\V{x} \Box\psiR{x}
+\psibarR{x}\Box(\Vdagger{x} \psiL{x})]\nn
\eea
along with the usual actions for the gauge and scalar fields,
where $\psiR{}$ and $\psiL{}$ are right- and left-handed chiral fermion fields
$\half(1-\gamma_5)\psi$ and $\half(1+\gamma_5)\psi$ respectively,
$\Dmu$
and $\Dtildemu$ are the forward and backward lattice derivatives
respectively, gauged with respect to the gauge field for the left chiral
symmetry $\ULmu{}$, $\Box$ is the lattice laplacian,
and $\V{}$ is a scalar field with frozen radius taking its value
in a group $G_L$ which for our purposes is taken to be $U(1)$ or $SU(2)$.

Aside from the term `Wilson--Yukawa' term containing $w$, this action is just
that which produces a naive (doubled) theory of gauged chiral fermions
coupled to a scalar field via a Yukawa interaction. The only differences
between this and the usual standard model are the presence of the doublers,
and an extra neutrino with right-handed chiral symmetry.
In the broken phase at tree level
the Wilson--Yukawa term takes the form $wv\psibar\Box\psi$ where $v =
\langle\V{}\rangle$, so that naively, this term looks like a Wilson mass
term, which could serve as a remedy for the doubling problem.

The action possesses a local chiral gauge symmetry
\bea
\psiL{x}&\to&\g{x}\psiL{x}\nn\\
\V{x}&\to&\g{x}\V{x}\\
\ULmu{x}&\to&\g{x}\ULmu{x}\gdagger{x+\muhat}\nn
\eea
which we will refer to as $g$-symmetry.

The phase diagram for this model for small gauge coupling
is well known by now \cite{Phasediagram} and is represented in figure 2.
\begin{figure}[t]
\epsfxsize=\columnwidth
\epsffile{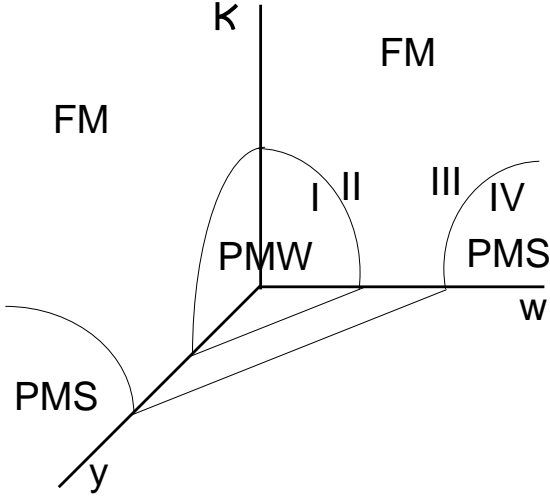}
\caption{the phase diagram for the Smit--Swift model at small gauge
coupling}
\end{figure}
The pure Higgs model exhibits a
phase transition for some $\kappa_c$, below which lies a symmetric phase and
above a broken phase. Taking $y\to0$ insures the presence of at least one
light fermion\cite{Usdecouple}, so we will limit our discussion to that case.
In the weak $w$ region,
the critical point extends into the $w-\kappa$ plane, separating a
symmetric (PMW) phase from a broken (FM) phase.
Perturbation theory applies in this region, and
just as in usual standard model perturbation theory, all fermion masses are
proportional to the vacuum expectation value of the Higgs particle
$\langle V\rangle=v$, including the masses of the doublers:
\be
m_f = v(y+2nw)\;\;\;n=0,1,2,3,4.
\ee
Therefore, in the PMW phase (region I
of figure 2), all $16$ fermions are massless, and in the
broken phase (region II), the degeneracy is lifted due to a non-vanishing
value of $v$. However, as the continuum limit is approached, $v$ scales
to zero, and all doublers appear in the physical spectrum. Thus the weak
coupling region is clearly not a region in which to look for a solution to
the doubler problem. This has also been checked via Monte Carlo
calculations\cite{BockpointA}.

To study the strong $w$ region, it is convenient to change variables to
fermion fields, neutral with respect to $g$-gauge symmetry:
\be
\psin{x} = \Vdagger{x}\psi_{x},
\ee
in terms of this field, the fermion part of the action becomes
\bea \SpWY &=
\sum_{x,\mu}&[\half\psibarLn{x}\gamma_\mu(\DWmu+\DWtildemu)\psiLn{x} \nn\\
&+&\half\psibarRn{x}\gamma_\mu(\latdmu+\latdtildemu)\psiRn{x}] \\
&+&\sum_x \psibarn{x}(y-\frac{w}{2}\Box)\psin{x},\label{eq:Neutralaction}\nn
\eea
where $\DWmu$
is a covariant derivative with respect to the composite gauge
field $\Wmu{}=\Vdagger{x}\Umu{x}\V{x+\muhat}$:
\be
\DWmu\psiLn{x}=\Wmu{x}\psiLn{x+\muhat}-\psiLn{x}.
\ee
We should note that this action looks like that of a chiral gauge theory
invariant under the following symmetry \cite{MaianiRome,UsRome},
(which we will here denote as $h$-symmetry):
\bea
\psiLn{x}&\to&\h{x}\psiLn{x},\\
\Wmu{x}&\to&\h{x}\Wmu{x}\hdagger{x+\muhat},\nn
\label{eq:hsymmetry}
\eea
along with a symmetry breaking Wilson mass term.
This form of the lagrangian appears as the starting point for
the Rome proposal with the relation being made manifest in unitary
gauge ($\V{}=1$), in which $\psi{}$ and $\psin{}$ are
equivalent as are $\Umu{}$ and $\Wmu{}$. We will return to this point later.
(See also \cite{FunKasAoki,SmitSeillac}.)

At strong $w$, the Schwinger--Dyson equations for the
right-handed fermion, which
happen to correspond to the Ward identities for a rigid fermion shift
symmetry, lead to some conclusions about the effective action of the
model\cite{JanandUs}. In particular,
$\psiRn{}$ does not appear in any interactions,
and the fermion mass spectrum is
\be
m_f^{(n)}=\sqrt{Z_2}(y+2nw),\;\;\;n=0,1,2,3,4,
\ee
where $Z_2$ is the wave function renormalization
which can be determined by expansion techniques or by Monte Carlo
calculations.  The important thing is that it is finite and non-vanishing.
Thus in the broken (FM) phase all fermions are massive, and a
single massless fermion can be arranged by tuning $y$ to zero. In the
symmetric phase the spectrum is exactly the same.

In addition, an examination of the propagator for the composite field
$\psic{x}=\V{x}\psin{x}$ which has non-zero charge with respect to
$g$-symmetry reveals that in the absence of gauge fields, apparently no
fermion states with non-singlet $g$-quantum numbers appear in the symmetric
phase\cite{BockJapan,Bockcharged,oneoverw,Aokilat92}. One can understand this
by
realizing that the large coupling $w$ has caused the original left-handed
charged fermion $\psiL{}$ to form a bound state with $\Vdagger{}$ and this
in turn pairs up with $\psiR{}$ to form a Dirac fermion, screening the
charge of the original state.  In the broken phase, $\psic{}$ mixes with
$\psin{}$ and does not represent an indepedent state.

With only the neutral fermions as physical states, we next must examine the
couplings to these fermions to see if they interact. We have already noted
that in the effective action, all
bare interactions of the right-handed fermion vanish. For the left-handed
fermion, it is a simple matter to see that its Yukawa interaction with the
Higgs field is proportional to $v$
which vanishes in the PMS phase and scales to zero in the FM phase near the
phase transition (region III).
The situation is similar for higher order
couplings between this fermion field and the scalar. Finally,
for small gauge coupling and hopping parameter $\alpha=1/(4w+y)$,
one can show that the gauge coupling to the neutral field also
vanishes\footnote{One author is not in agreement\cite{Aokicouplings},
but I believe the arguments referred to here to be correct.} in
the scaling region\cite{JanandUs,UsJapan},
and this can be demonstrated in the broken phase
taking into account the contributions
of the scalar field exactly\cite{UsRome}. The
interpretation is that once the scalar fields form bound states
with the fermions so as to form Dirac fermions which are neutral with
respect to the $g$-symmetry, they then screen these fermions from any
interaction with the gauge fields.

A quick way to look at this is to notice that in the interaction lagrangian,
$\Wmu{}$ couples to the fermions through the operator
$\Vdagger{x}\Dmu\V{x}-\hc$ which is a dimension three operator in contrast
to the usual gauge field. Thus in the continuum limit the interaction term
is dimension $6$ and will need two extra powers of the lattice spacing to
compensate. In the broken phase this means that the interaction must be
proportional to $v^2\equiv{\overline v}^2a^2$ (where ${\overline v}$ is
dimension $1$), and in the symmetric phase a factor of $\Lambda^2a^2$ will
emerge, where $\Lambda$ is the scale associated with the gauge fields (e.g.
``$\Lambda_{\rm QCD}$''). In both cases, the interaction vanishes in the
continuum limit.

Now we can summarized what we learned so far. First,
fermions pair up as vectors everywhere in the phase diagram. Second,
There are no $g$-charged fermions in the PMS phase. And third,
in the strong $w$  region, the $g$-neutral ($h$-charged) fermion
is a free particle.  It is possible to define a theory with
$g$-charged states that appear in the physical spectrum, by changing the
form of the Wilson--Yukawa term, and these do couple
with the gauge field, but they do so in a vector-like manner\cite{JanandUs}.
Hence we have learned several valuable
lessons from the study of this model which
may serve as warnings for future investigations:
\begin{itemize}
\item If there is a strong coupling in the game, and chirally opposite bound
states can form to pair up as Dirac fermions, they likely will.
\item In this model, the scalar field $\V{}$ is the culprit. $\V{}$ screens the
chiral gauge interactions and decoupling $\V{}$ by letting its mass diverge (as
deep in the PMS phase) doesn't work.
\item It is not a priori clear which states are physical by looking at the
lagrangian.
\item Anomalies play no role (see also next section)
\end{itemize}
One final comment: note the dimension dependence of the conclusions
regarding the gauge interaction term. In two dimensions, the operator
$\Vdagger{x}\Dmu\V{x}-\hc$ is dimension one and the quick argument above
would not require the interaction to vanish. Indeed there is evidence that
in a two dimensional theory one can construct a theory of chiral fermions
using the Smit--Swift model\cite{Bock2d}. This case also gives us the first
scenario for escaping the Banks ``$U(1)$ problem''. In two dimensions,
apparently the fermion forms a bound state with the scalar to form a vector
multiplet, but the left-handed component nevertheless interacts with the
composite gauge field in a chiral fashion. Thus $h$-symmetry emerges as the
symmetry associated with this coupling. This turns out to be a realization
of the Dugan-Manohar solution to this problem\cite{DuganMan}, namely that
the current associated with the relevant gauge symmetry is not what one
would expect a priori. Indeed the $U(1)$ rigid symmetry commutes with
$g$-symmetry, but the states are classified according to $h$-symmetry. The
current for this symmetry is related to that for $g$-symmetry by a local
counterterm, and so the Noether current for the rigid $U(1)$ symmetry is not
the $h$-gauge invariant current. In other words, the rigid $U(1)$ symmetry
and the $h$-symmetry do not commute. Although the Dugan--Manohar scenario is
not realized in four dimensions due to the screening of the interaction, it
still teaches us that the way to the continuum can be more subtle than first
imagined.

\subsection{The Eichten--Preskill Model}

Eichten and Preskill assumed from the start that when constructing a
lattice theory with any hope of defining a theory of chiral fermions,
one must pay careful attention to anomalies\cite{EichPres}. Therefore
they proposed to look
at a theory that respects all desired symmetries of the target continuum
theory, and explicitly breaks any symmetry that should be broken. To satisfy
this criterion, they chose to study a lattice version of a chiral $SU(5)$
grand unified theory.

For simplicity and with no loss I will discuss an $SO(10)$ theory with the
gauge fields turned off. The $SU(5)$ theory can be obtained by adding
explicit symmetry breaking terms to the action.
The action for the $SO(10)$ theory can be written as follows.
\bea
S_{\rm EP}&=&\sum_{x\mu}\half[\psidaggerL{x}\sigmahatmu\psiL{x+\muhat}-\hc]\\
&-&\sum_x\frac{\lambda}{24}[(\psitransposeL{} \sigma_2 T\psiL{})^2+\hc]\nn\\
&-&\sum_x\frac{r}{48}[\Delta(\psitransposeL{} \sigma_2 T\psiL{})^2+\hc],\nn
\eea
where $\psiLi{}$ is a Weyl spinor in the $16$ representation of $SO(10)$,
$T^{aij}$ is an $SO(10)$ invariant tensor transforming as ${\overline
16}\times{\overline 16}\times 10$, $\sigmahatmu=(1,i{\vec\sigma})$, where
$\vec\sigma$ are the Pauli matrices, and
\bea
\Delta(\psi_i\psi_j\psi_k\psi_l)&=&-\half\sum_{\pm\mu}
[\psi_{i x+\muhat}\psi_{j x}\psi_{k x}\psi_{x l}\\
& +&\cdots-4\psi_{i x}\psi_{j x}\psi_{k x}\psi_{x l}].\nn
\eea
This latter construction allows the term in the action proportional to $r$
to play the role of a Wilson mass term, breaking the degeneracy in the mass
spectrum between the fermion and its doublers.  Note that the quantity
$\psitransposeL{}\sigma_2T^a\psiL{}$
explicitly breaks fermion number and so by construction the
Eichten--Preskill model does not have the Banks problem.

Studying the model for $r=0$, Eichten and Preskill found a symmetric phase
in the strong $\lambda$ region with massive fermions, and another symmetric
phase in the weak coupling region but with massless fermions. They then
hoped that at the phase transition between them, the degeneracy of the
fermion and its doublers would be lifted by turning on the coupling $r$, and
that a region would exist where the one fermion can be tuned to become
massless, whereas the doublers have masses of the order of the cutoff. They
also pointed out that such a scenario would not be realized if the two
symmetric phases were separated by a broken phase.

The model above is rather like the Nambu--Jona-Lasinio model in appearance,
and recent work showing the equivalence of a model of
this type with the standard model\cite{Kutietal} suggests that a similar
association can be made with the Eichten--Preskill model. Indeed this is the
case. An action with the same symmetries as the model in question
is
\bea
S^\prime_{\rm EP}
&=& \sum_{x\mu}\half[\psidaggerL{x}\sigmahatmu\psiL{x+\muhat}-\hc]\\
&+&\half\sum_x\phi_x\phi_x -\kappa\sum_{x\mu}\phi_x\phi_{x+\muhat}\nn\\
&+&\half y\sum_x\psitransposeL{} \sigma_2 T\psiL{}\phi+\hc\nn\\
&-&\frac{1}{4} w\sum_x\psitransposeL{}\sigma_2T\Box\psiL{}\phi+\hc\nn
\eea
where we have introduced a scalar $\phi^a$ in the $10$ representation of
$SO(10)$. This model with Wilson--Yukawa term
obviously has a similar structure as the Smit--Swift
model. To make contact with the Eichten--Preskill model,
one can invent an additional `flavor' and
perform a large $N$ expansion in the number of flavors.
To do this, we
studied the Weyl fermion $\psiL{}$ (explicit in both models), its
right-handed partner $\psiR{}$ (formed as a bound state in both models), and
the scalar field (a bound state in the Eichten--Preskill model), whose
composite forms are
\bea
\hbox{\rm field}&\hbox{\rm Eichten--Preskill}&\hbox{Wilson--Yukawa}\nn\\
\psiL{}&\psiL{}&\psiL{}\nn\\
\psiR{}&\sigma_2T^{\dagger a}\psistarL{}
(\psidaggerL{}\sigma_2T^{\dagger a}\psistarL{})&
\phi^a\sigma_2T^{\dagger a}\psistarL{}\\
\phi^a&{\rm Re}\psitransposeL{}\sigma_2T^a\psiL{}&\phi^a\nn
\eea
where a sum over $a$ is implied when repeated.
Not surprisingly, the phase diagram of the above Wilson--Yukawa model is
similar to that of the Smit--Swift model (figure 2), with the
parameters $(y,w)$ playing the role of
$(\lambda,r)$ in the Eichten--Preskill model.
One can find matching conditions between the couplings of the two models so
that all Green functions involving the particles above coincide in the large
$N$ expansion\cite{UsEichPres}.
The physics of the Eichten--Preskill model is realized in the $\kappa=0$
plane, and in particular, the two symmetric phases are separated by a broken
phase. In the strong $\lambda$ region, the right-handed fermion forms as a
bound state and pairs with the original fermion to give a theory of vector
fermions interacting with gauge fields, and in the weak coupling region the
doublers are present in the physical spectrum.  Thus all that we learned about
the Smit--Swift model applies here, and again no theory of chiral fermions
is realized.  Finally we should emphasize that even though Eichten and
Preskill were careful to pay attention to the anomaly structure of the
theory, this plays no role in its failure.

\subsection{The Rome proposal}

An understanding of the preceding problems, and in particular that the
culprit for the failure of the Smit--Swift model to produce a theory of
chiral fermions is the scalar field forming bound states with the fermions
to create Dirac partners, suggests a direction for proceeding.  The scalar
field represents just those gauge degrees of freedom which are unphysical,
and would be removed by gauge fixing. This is easily seen in a transverse
gauge for which the scalar fields represent the longitudinal modes.  Since
we have learned that
decoupling of these modes by keeping their mass at the cutoff does
not prevent the bound states from being formed and destroying the chiral
structure of the theory, perhaps enforcing gauge
fixing on the lattice would do the trick. This is precisely
the proposal of the Rome group\cite{Roma}, although historically this
proposal was quite independent from the reasoning I have presented here.
Originally the motivation of the Rome group was to define a theory motivated
by perturbative gauge fixing as a prescription for obtaining a full
non-perturbative asymptotically free chiral gauge theory.

The action for the Rome group's proposal starts with that given in
\eq{Neutralaction}, but without the scalar field (as it appears in
unitary gauge).  Then they add gauge fixing and ghost terms of the form
required for a particular gauge choice such as Landau gauge. Thus they
start with a theory that violates gauge invariance (here we are talking
about $h$-gauge invariance, not involving the scalar fields), and the final
ingredient is to add all counterterms necessary to allow tuning to impose
the satisfying of the BRST identities associated with $h$-symmetry:
\be
\SRoma = \SpWY(V=1)+\Sgf+\Sghosts+\Sct.
\ee
With
this procedure the longitudinal modes are decoupled not by putting their
mass at the cutoff scale, but by decoupling their interaction with other
particles of the theory.

Can his decoupling be accomplished? It appears that the prescription should
work in principle to all orders in perturbation theory, and indeed there are
now some explicit two loop results indicating that things are working
at least to that order\cite{Rometwoloop}.
The question arises how well it can work as a non-perturbative
prescription, in which the tuning must (at least in part) be done
numerically. To address this question, let us take a look at the path
integral of the theory. The path integral for the Rome proposal is given by
\be
\int [dU_\mu][d\psi][d\psibar]\;e^{-\SRoma}.
\ee
One
procedure to approach this path integral is to multiply by the trivial integral
$\int [dV]=1$, and then make a transformation of variables to rotate $V$
back into the action\cite{FunKasAoki,SmitSeillac}, so that the first term looks
like $\SWY$ as given in
\eq{SSaction}. As Smit has pointed out
(see e.g. \cite{SmitRome}) the longitudinal modes are actually present in
the Haar measure for the gauge fields, and the above trick is merely making
explicit what already is there. Thus in so far as the Rome proposal would
work, it is obvious that it depends on how well they are able to decouple
the scalars.
If the decoupling of the scalars is very sensitive to tuning then values of
the coupling that are a little off may not prevent the scalars from forming
bound states with the fermions and we would be back to the scenario of the
Smit--Swift model. Because it is expensive to tune couplings numerically,
this is a potentially serious problem. As to whether the problem is
realized, ``the proof is in the pudding'' so to speak, and we will have to
wait for a realistic attempt at doing the non-perturbative problem.

Bodwin and Kovacs\cite{BodKov} have made an observation that could make the
problem more manageable, provided the technical difficulties can be worked
out. They observe that the magnitude of the chiral fermion determinant is
equal to the positive square root of a vector determinant, and provided a
method is found to deal with the phase of the chiral determinant, the
calculation could be done with a vector theory plus an extra calculation for
the phase. In the abelian case this reduces the number of counterterms
needed from
seven down to two, which is a substantial savings, and similarly in the
non-abelian theory. There is also a proposal
on the market for calculating the phase\cite{Gockeler} using topological
methods following a definition by Alvarez-Gaum\'e et. al.\cite{AlvDP}, so a
combination of all three of these ideas may eventually prove fruitful, if
not elegant.

Finally, I would like to point out how the Rome approach gets around the
Banks problem. Actually so far two ways have been proposed\cite{MaianiRome}.
One way around it is to write each chiral multiplet as a charged chiral
fermion and a neutral `spectator' fermion, in which the neutral spectator
fermion obeys shift symmetry. Then using the decoupling
theorem\cite{Usdecouple} the spectator fermions should all decouple, and the
$S$-matrix should factorize into charged particles and spectators.
Kikukawa has already shown that the combination
of a right and left charged fermion along with the respective spectators
can produce the correct anomaly
responsible for baryon number violation\cite{Kikukawa}. So despite the fact
that the full $S$-matrix including spectators would be $U(1)$ invariant, due
to the decoupling of the spectators, the baryon number would be carried off by
the spectators.
The second method is to use Wilson-Yukawa terms that have the form of
Majorana mass terms\cite{RomeMaj}.
These explicitly violate the $U(1)$ symmetry, and Pryor
has shown (in the context of the Smit-Swift type models) that such a term
reproduces the correct anomaly\cite{Pryor}.

\section{Staggered fermion approach}

With the demise of the Wilson--Yukawa approach, this past year has also seen
renewed interest in the staggered fermion\cite{Staggered,KawaSmit,STWeisz}
approach toward constructing a
lattice standard model\cite{SmitSeillac,SmitRome,StaggeredMC}.
For details of the method, see \cite{SmitRome,definestaggered}.
First, the basic idea of the staggered fermion
theory is to spread out the spin and flavor
components of a fermion on the lattice,
resulting in the decoupling of the original $16$ doubled fermion
components into $4$ independent $4$-plets. Then only
one of these
$4$-plets is kept to formulate the theory, thus reducing the $16$ original
flavors to $4$. Because of this spreading out of the spin components,
the hypercubic symmetry and flavor symmetry are now mixed,
and thus for example, a shift on the lattice of one lattice spacing
mixes flavors, whereas it takes two shifts to generate a
spatial translation. Also $\gamma_5$ corresponds to a four link operator.
Now the beauty of the staggered fermion approach is that rather than trying
to get rid of the remaining doublers, they are to be used as physical
flavors in the theory. The
important consequence of this is that if the four flavors are to correspond
to flavors in the continuum theory, since their components
are sitting on different
sites of the lattice, the global chiral symmetry is broken
and therefore cannot be gauged in an invariant way. (QCD escapes this by
adding an independent color index to each flavor, and the new symmetry is
gauged.) This is reminiscent of the Rome proposal in which chiral gauge
invariance is broken explicitly on the level of the action and must be
restored by tuning, and so it is with staggered fermions.

It is useful to keep track of
two remnant $U(1)$ symmetries that are not broken, one
transforming each fermion component by the same phase,
relating to fermion number and the second rotating alternate components by
opposite phases:
\be
\exp(i\omega\epsilon_x)\in U_\epsilon(1), \;\;\;
\epsilon_x = (-1)^{x_1+x_2+x_3+x_4}.
\ee
These are of course part of a rich lattice symmetry group including many
other discrete symmetries which I will not enumerate here. The group goes
under the name $SF$ or staggered fermion symmetry group\cite{definestaggered}.

If $\chi$ is the component of a fermion at a particular site, one can cut
the number of components in half by letting $\chibar=\chi$ on each site
(equivalent to letting $\chi$ live only on even sites and $\chibar$ on odd
sites). These are referred to as reduced staggered fermions. Each reduced
fermion field represents a doublet of fermion flavors, and because of the
`Majorana' constraint that defines them, the $U(1)$ relating to fermion
number is broken and only $U_\epsilon(1)$ is left.

It is straightforward now to build a lattice theory with particle content
equivalent to the standard model. For example\cite{SmitRome},
for the first generation of
the standard model, the electron and neutrino together can be represented by
one $\chi$ doublet, and the three colors of $u$ and $d$ quarks can be
represented by three fields $\chi_a$ with the color index $a=1,2,3$. (In
this formulation, the $U_\epsilon(1)$ symmetry may lead to fermion number
conservation in the scaling limit and presumably
the Banks problem arises. A somewhat less elegant embedding of standard
model quantum numbers into staggered fermion doublets can be made however
that explicitly breaks $U_\epsilon(1)$ and avoids the
problem\cite{SmitRome}.)

To gauge the model, a more useful organization of the fermion components is
to define the Grassmann matrices
\bea
\Psi{x}&=&\frac{1}{8}\sum_b\Gamma(x,b)\half(1-\epsilon_{x+b})\chi_{x+b},\\
\Psibar{x}&=&\frac{1}{8}\sum_b\half\Gamma(x,b)(1+\epsilon_{x+b})\chi_{x+b},\\
\Gamma(x,b)&=&\gamma_1^{x_1+b_1}
\gamma_2^{x_2+b_2}\gamma_3^{x_3+b_3}\gamma_4^{x_4+b_4}
\label{eq:Psidef}
\eea
where $b$ is summed over all corners of a unit lattice cell.  In terms of
these fields one can define a chiral model in a notation reminiscent of that
of the continuum theory. For example, and $SU(2)\times SU(2)$  model in
which only $SU(2)\times U(1)$ is gauged, can be defined through the action
\bea
S=&-&\sum_{x\mu}\half\tr[\Psibar_{x}\gamma_\mu\Psi_{x+\muhat}\Udaggermu{x}\nn\\
  &-&\Psibar_{x+\muhat}\gamma_\mu\Psi_{x}\Umu{x}]\nn\\
  &+&\sum_{x\mu}m_\mu\rho_x\tr[\Psibar_{x}\Psi_{x}\gamma_\mu],
\eea
where the gauge fields are the appropriate ones for gauging the groups
desired and the last term is present since no symmetry excludes it.
$\rho_x$ is a spacetime dependent amplitude factor.
Keep in mind that because the components of $\Psi$ lie on different sites,
this action is not gauge invariant. However, if one sets
$m_\mu$ to zero, the action does lead to the proper gauge invariant
continuum action in the classical continuum limit.

Now recall that in the Smit-Swift model one could freely transform between
the $g$-charged fields $\psic{}$ and the $g$-neutral fields $\psin{}$, and
represent the path integral in terms of either. Such a transformation is not
possible here because of the lack of local gauge invariance. One can however
expose the longitudinal degrees of freedom by making the transformation of
variables $\Umu{x}\to\V{x}\Umu{x}\Vdagger{x+\muhat}$\cite{SmitSeillac}, then
one arrives at the action ($m_\mu=0$)
\bea
S=&-\sum_{x\mu}&\half\tr[\Psibar_{x}\gamma_\mu\Psi_{x+\muhat}
\V{x+\muhat}\Udaggermu{x}\Vdagger{x}\\
       &&-\Psibar_{x+\muhat}\gamma_\mu\Psi_{x}\V{x}\Umu{x}\Vdagger{x+\muhat}]
\label{eq:staggeredneutral}
\eea
which appears similar in form to the version of the Smit--Swift model
written in terms of neutral fields in \eq{Neutralaction}.
The obvious question based on our
previous experience is whether the corresponding desired
`charged' states appear in
the physical spectrum, or equivalently, whether the propagator
\be
\langle\Vdagger{x}\Psi_{x}\Psibar_{y}\V{y}\rangle
\ee
has a pole structure (when the gauge fields are turned off). This is a
crucial test as to whether a gauge theory emerges which includes charged
states in the usual sense.

Alternatively, one can define another fermion--Higgs model through an action
which has the form of \eq{SSaction}
\bea
S=&-&\sum_{x\mu}\half\tr[\Psibar_{x}\gamma_\mu\Psi_{x+\muhat}
        -\Psibar_{x+\muhat}\gamma_\mu\Psi_{x}]\nn\\
  &+&y\sum_{x}\tr\Psibar_{x}\Psi_{x}\Phi_{x}+S_{\rm scalar},
  \label{eq:staggeredcharged}
\eea
where $\Phi=\sum_\mu\Phimu{x}\gamma_\mu$. Because there is no formal
transformation between the models defined by
\eq{staggeredneutral} and by \eq{staggeredcharged}
it is a non-trivial question whether the two quantum theories are
equivalent.  Hence it makes sense to first study the two fermion--Higgs systems
alone, for if these are seen to be equivalent, one might expect the
gauged theories to be also.  Then the more pressing question can be
addressed: whether they (one or the other or both) give rise to the desired
gauge invariant continuum theory (assuming the necessary tuning is performed).

As a first serious investigation of these questions, recently a group has
begun to study an $SU(2)\times SU(2)$ invariant version of
the latter model numerically\cite{StaggeredMC}. I will
summarize the highlights here as befit the flow of my
theme; a more complete report can be found
elsewhere in this volume\cite{Bocklat92}.
To define the model, the action for the scalar field
\bea
\lefteqn{S_{\rm scalar}=\kappa\half\sum_{x\mu}
\tr(\Phidagger_{x}\Phi_{x+\muhat}+\Phidagger_{x+\muhat}\Phi_{x})}\\
&&-\half\sum_x\tr[\Phidagger_{x}\Phi_{x}+\lambda(\Phidagger_{x}\Phi_{x})^2],
\eea
is added to \eq{staggeredcharged}.  The case
of $2$ reduced staggered doublets or $4$ flavors
and the case of $2$ multiplets of naive fermion corresponding to $32$
flavors were studied.

First of all, the phase diagram has been mapped out and no surprises were
found. For weak values of $y$ and large $\kappa$ a ferromagnetic phase
occurs. As one reduces $\kappa$ a second order transition appears to a
paramagnetic phase which continues on into the negative $\kappa$ region,
after which for large enough negative $\kappa$ an antiferromagnetic phase
occurs. For much larger values of $y$ the diagram moves directly from the
ferromagnetic to a ferrimagnetic phase for some value of $\kappa$ below
zero. As a by-product in the simulation it was discovered that the staggered
fermion method appears to offer a very efficient way to simulate such
models.

Another question addressed is $O(4)$ symmetry restoration. Because of the
explicit breaking of the symmetry, two counterterms $O^{(1)}$ and $O^{(2)}$:
\be
O^{(1)}=\sum_{x\mu}\Phi^4_{\mu x},\;\;
O^{(2)}=\sum_{x\mu}(\Phi_{\mu x+\muhat}-\Phi_{\mu x})^2,
\ee
are necessary to restore it as the continuum is approached, which in
principle would mean that two couplings need to be tuned. What was found
however, is that in the region studied, the $O(4)$ symmetry is already
present to a good approximation, and that the breaking effects are only on
the order of a few per cent. Two other facets of the calculation deserve
mention before going on. The authors have included one loop fermion effects
in fitting the scalar parameters, and in order to control systematic errors
due to finite size effects, the authors have performed calculations on
lattices of different physical sizes, in order to make an extrapolation to
large volume.
Among results obtained from these models are estimates of the effects of
fermions on the upper bound of the Higgs mass, which brings me to the next
topic.

\subsection{Higgs mass upper bound revisited}

We are now back to the topic I mentioned earlier in my talk in
the context of pure scalar theory. Estimates of the fermion effects
on the Higgs mass upper
bound have also been obtained using mirror fermions\cite{Montvayetal}.
Figure 3 is a summary of these various
results for the upper bound of the Higgs mass.
\begin{figure}[t]
\epsfxsize=\columnwidth
\advance\epsfxsize by -0.5cm
\setbox\rotbox=\vbox{\epsffile{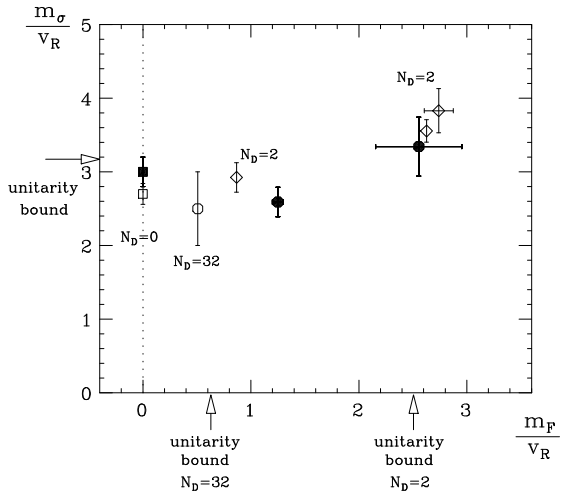}}\rotl\rotbox
\caption{Estimates for Higgs mass upper bound in pure Higgs theory and with
fermions. The standard action pure Higgs theory on a hypercubic lattice is
given by the white box, and the results of Heller et. al. are the black box.
The circles are those including mirror fermions, and the diamonds are with
staggered fermions.}
\end{figure}
This figure is a plot of the renormalized mass of the Higgs particle,
$m_\sigma$, versus the renormalized mass of
the fermion, $m_f$, both in units of the
renormalized Higgs vacuum expectation value, $v_R$ ($\equiv f_\pi$ from
section 2).
The data comes largely from refs. \cite{Heller},\cite{Montvayetal}
and \cite{StaggeredMC}.  $m_f=0$ represents the case without fermions. The
white box represents the data from previous calculations on a hypercubic
lattice with the standard action, and the black box is the number quoted
from Heller et. al. \cite{Heller}
earlier in the talk.  The circles are estimates from the mirror fermion
method, and the diamonds come from the infinite volume extrapolation of the
calculations with the staggered fermion method.
For details of how the numbers were obtained, please
see the appropriate references.
First of all, notice that the bounds are in fairly good agreement between
the two fermion methods. This is reassuring and is a check that the
regularization dependent affects are not dominant. In fact for rather large
fermion mass, these results are close to the tree level unitarity bound, which
indicates that the renormalized couplings are still rather small.
Secondly, note that the inclusion of fermions allows a
slightly higher (perhaps 30\%) upper bound estimate
than that found from the pure Higgs sector, perhaps as high as 1 TeV.
This, accompanied by the broadening of the decay width,
is an indication that finding the Higgs particle
might be more difficult than we thought.

\section{Domain wall fermions}

A new and perhaps the most interesting proposal presently on the market for
putting chiral fermions on the lattice, goes off in a different direction
(quite literally). Based on work by Callan and Harvey\cite{CalHarv},
showing that a vector
fermion theory in $2n+1$ dimensions in the presence of a domain wall scalar
field, admits a $2n$ dimensional chiral fermion
solution, Kaplan suggests that this approach might be used in the lattice
theory as well\cite{Kaplan}.  The beauty is that because the theory is
vector-like in $2n+1$ dimensions, A Wilson term can be used to make the
doublers massive. So we start from the standard Wilson action for a
free fermion in $5$
dimensions (I will limit myself to $n=2$ in what follows) with the exception
that the `mass' of the fermion is actually a background scalar field in a
domain wall configuration:
\bea
S&=&\sum_x\half\sum_{\mu=1}^5\psibar(x)\gamma_\mu(\latdmu+\latdtildemu)
\psi(x)\\
&+&\sum_x\psibar(x)(m(x)+\frac{r}{2}\Box)\psi(x),
\label{eq:domainwallaction}
\eea
where
\be
m(x) = \left\{ \ba{rl} m&x_5>0\\0&x_5=0\\-m&x_5<0\ea \right.
\ee
in which $m>0$, and $\Box$ is now the five dimensional lattice laplacian.

Kaplan shows that in this theory, a normalizable
solution of the associated Dirac equation
exists which is a simultaneously a chiral zero mode of the
four dimensional lattice Weyl equation on the domain wall, and no other
normalizable zero modes exist. The argument goes as follows. The Dirac
equation following from \eq{domainwallaction} is given by (for $r=1$)
\bea
\lefteqn{\pr+5\psi(p,x_5+1)+\pr-5\psi(p,x_5-1)}\\
\lefteqn{\;\;+\meff(p,x_5)\psi(p,x_5)
+i\sum_{\mu=1}^4\gamma_\mu \sin p_\mu\psi(p,x_5) = 0,}\nn
\eea
with
\bea
\meff(p,x_5)&=&m(x_5)-1-F(p),\\
F(p)&=&\sum_{\mu=1}^4(1-\cos p_\mu),
\eea
and $\pr\pm\mu=\half(1\pm\gamma_\mu)$.
Now if we take the fourier transform in four dimensions, then solutions
of the form ($p$ is the $4$-momentum)
\be
\psi_\pm(p,x_5) = e^{ip\cdot\xtilde}\phi_\pm(p,x_5)
\ee
satisfy the four dimensional lattice Weyl equation
\be
i\sum_{\mu=1}^4\gamma_\mu \sin p_\mu \psi_\pm = 0,
\ee
if $\phi_\pm$ satisfy
\bea
\phi_+(p,x_5+1)+\meff(p,x_5)\phi_+(p,x_5)=0,\\
\phi_-(p,x_5-1)+\meff(p,x_5)\phi_-(p,x_5)=0.
\eea
Solutions are easily obtained by assuming a value for $\phi(p,0)$ and
hopping. The solutions either grow exponentially with $x_5$ or decay
according to the size of $M(p,x_5)$. Normalizable, left-handed (positive
chirality) solutions exist if
\bea
&|m-1-F(p)|<1, & x_5>0,\\
&|m+1+F(p)|>1, & x_5<0.
\label{eq:leftchiralcond}
\eea
and the conditions for right-handed (negative chirality) normalizable
solutions are similar, but with the inequalities reversed:
\bea
&|m-1-F(p)|>1, & x_5>0,\\
&|m+1+F(p)|<1, & x_5<0.
\label{eq:rightchiralcond}
\eea
If $m<2$, the conditions for a left-handed chiral solution
are satisfied if $0<m-F(p)<2$ or $F(p)\approx0$, and
one left-handed chiral solution exists.  The conditions for a right-handed
chiral fermion on the other hand cannot be satisfied.  For other values of
$m$ the situation changes somewhat: for the ranges $2<m<4$,
$4<m<6$, $6<m<8$ and $8<m<10$ there are four right-handed solutions, six
left-handed solutions, four right-handed solutions and one left-handed
solution respectively, with no solutions for larger values of
$m$\cite{SchmaltzJansen}.

To see what has happened to the usual doublers of chiral fermions we should
put the theory in a box of size $L$ in the fifth direction, and impose
periodic boundary conditions. Then the domain wall mass becomes
\be
m(x) = \left\{ \ba{rl} 0&x_5=0\\m&0<x_5<L/2\\
		0&x_5=L/2\\-m&L/2<x_5<L\equiv0,\ea \right.
\ee
so that there are now two domain walls. Following similar arguments as
above, one finds that for every chiral fermion bound to one domain wall
there  is one of opposite chirality bound to the other. Thus the doubling is
still there, but the chiral fermions are separated by a distance $L/2$ in
the fifth dimension, and communication between them is damped exponentially by
the mass of the vector particles of the theory. Thus if $m$ is taken large (as
the cutoff), the four dimensional hyperplanes decouple and we have two
mirror worlds of free chiral fermions which don't interact. To summarize,
for $m<2$ say, we have a spectrum of one chiral fermion on each domain wall,
and otherwise Dirac fermions with mass $m$ and doublers with a mass of the
order of the cutoff.

Now that we have defined a theory of free chiral fermions, the central
question is whether they can be coupled to gauge fields.  The first test is
to study the theory with background gauge fields to see whether the correct
anomaly structure will be reproduce, as it is in the continuum case. In the
latter case, Callan and Harvey follow Goldstone and Wilczek\cite{GoldWil}
to obtain a
contribution to the fermion current which arises from the heavy fermions, as
$m$ is taken large.  From calculating the relevant triangle diagram they
find that
\be
\langle\psibar\gamma_\mu\psi\rangle=-\frac{i}{2}\frac{m(x)}{|m(x)|}
c\epsilon_{\mu\alpha\beta\gamma\delta}F_{\alpha\beta}F_{\gamma\delta}.
\ee
In particular this term survives as $m\to\infty$. Now the five dimensional
theory is exactly gauge invariant, so the five dimensional current is
conserved. For large $m$, the factor $m(x)/|m(x)|$ becomes a step function
so that
\bea
\sum_{\mu=1}^4\partial_\mu J^\mu &=& -\partial_5 J^5\\
&\propto&\delta(x_5)\epsilon_{5\alpha\beta\gamma\delta}F_{\alpha\beta}F_{\gamma\delta},
\eea
which is the correct anomalous divergence of a left-handed chiral fermion
in four dimensions.  Thus we see
that the anomaly can be interpreted as current flowing off the domain wall.
The effective
action whose variation is the anomaly can easily be calculated:
\bea
\Seff&\propto&\int d^5x\langle J_\mu\rangle A_\mu\\
&\propto&\int d^5x \frac{m(x)}{|m(x)|}A_\mu
\epsilon_{\mu\alpha\beta\gamma\delta}F_{\alpha\beta}F_{\gamma\delta},
\eea
which is proportional to the Chern--Simons density.  Thus the domain wall
mass has conspired with the gauge fields to give rise to a Chern--Simons
term.

The lattice version of the derivation follows along similar lines, although
it differs somewhat due to the richer fermion spectrum\cite{GolJanKap}.
%
%
%
The end result is that the factor $m(x)/m(x)$ must be replaced by the factor
\be
\sum_{n=0}^5(-1)^n\left(\ba{c} 5\\n\ea\right)\frac{m(x)-2n}{|m(x)-2n|},
\ee
which is consistent with the zero mode spectrum of ref.\ \cite{SchmaltzJansen}.
In addition, using the above ideas with regard to a three dimensional domain
wall theory, a direct numerical computation has been
performed of the anomalous Ward identity in two dimensions\cite{Jansen2d}.

So now we have a theory of chiral fermions that appears to work
when coupled to background gauge fields. The real test and the most
difficult task is to include the dynamics of the gauge fields. As we have
seen, this
has also been a problem with previous methods although for somewhat
different reasons. In
this case, we start with a five dimensional theory which somehow must be
made to look four dimensional in the end, so  in particular we would not expect
that a perturbation theory in all gauge fields could be performed due to the
non-renormalizability of the five dimensional
theory. Fortunately this is not necessary, for
five dimensional euclidean (Lorentz) invariance is already broken, so that
the gauge couplings in the four dimensions and the fifth can be taken
to differ:
\be
\Sgauge =\beta\sum_{i,j=1}^4\tr U_{ij}+\beta_5\sum_{i=1}^5\tr U_{i5},
\ee
where $U_{ij}(p)$ are plaquette variables.  Thus $\beta$ could be taken
large as usual for an asymptotically free chiral gauge theory, and $\beta_5$
could be varied,  as needed. Alternatively, we may consider a subset of the
possible five dimensional gauge fields, as long as gauge
invariance in the four
dimensional end product theory is maintained.

%

Another potential pitfall comes from the strange constraints relating
different regions in the Brillouin zone to the existence of chiral
solutions\cite{BodKovPriv}. These will cause the momentum space
propagators to be highly unusual,
and they could cause gauge invariance to be broken in the four dimensional
sense. Then in order to restore gauge invariance we would have to add
counterterms to the theory, bringing us right back to the scenario of the
Rome proposal. This brings one to ask, do all roads lead to Rome? One may
hope not, but in any case, the crucial test for this model remains to
understand how to add dynamical gauge fields.

A final note: assuming that somehow this model would survive the addition of
dynamical gauge fields and a lattice version of the standard model could be
defined. this model would also escape the Banks $U(1)$ problem, and perhaps
in the most novel way. Since only local anomalies need be canceled, the
global anomalies would remain present, and proton decay would be permitted by
sending baryon number off into the fifth dimension.

\section{TWO MORE PROPOSALS}

\subsection{Reflection positive fermions}

Another new proposal this year due to Zenkin
is an attempt to define a theory of
chiral fermions by enlarging  the Hilbert space and enforcing reflection
positivity\cite{Zenkin}. The method borrows a trick from constructive field
theorists\cite{Brydges}, writing the gauge field link variables $\Umu{x}$
as the product of two different group elements:
\be
\Umu{x} = \Wmu{x} W^\dagger_{-\mu x+\muhat},
\ee
so that each site in a $d$ dimensional theory is associated with
$2d$ $W$-fields, one for each direction both forward and backward, rather
than the usual $d$ link variables. Operators in the theory
are defined as functionals in this larger space, with the path integration
over
both $\Wmu{}$ and $W_{-\mu}$ independently. Finally, one can take each
$W_{\pm\mu}$ to have components that couple differently to right- and
left-handed fields,
\be
W_{\pm\mu x} = W^L_{\pm\mu x}\PL+ W^R_{\pm\mu x} \PR.
\ee
For a vector theory, there is no cost at enlarging the
Hilbert space in this way,
for a change of variables will remove half of the $W$-fields
from the action, and they can be integrated over trivially. This is not
true of a chiral theory.

The final step of the proposal is to construct an action that is reflection
positive, while using the Wilson mechanism to control the doublers. This
action has the form
\be
S = B + \Theta[B]+\sum_i C_i \Theta[C_i],
\ee
where $B$ and $C_i$ are functionals of fermions $\psi$ and gauge fields $W$
defined for $x_4>0$, and $\Theta$ is an antilinear operator such that
\bea
\lefteqn{\Theta[\psibar_x\cdots\Gamma\cdots W_{\pm\mu y}\cdots\psi_z]}\\
&&=\psibar_{r(z)}\gamma_0\cdots
W^\dagger_{\pm r(\mu) r(y)}\cdots\Gamma^\dagger\cdots\gamma_0 \psi_{r(z)},\nn
\eea
where $\Gamma$ is a matrix in spin space and $r$ denotes a reflection along the
$x_4$ axis. The action is
\bea
S&=\sum_{x\mu}&\half\psibar_{x}\gamma_\mu[
(\PL \ULmu{x}+\PR\URmu{x})\psi_{x+\muhat}\\
&&-(\PL \UdaggerLmu{x-\muhat}+\PR\UdaggerRmu{x-\muhat})\psi_{x-\muhat}]\nn\\
&-\sum_{x\mu}&\half\psibar_{x}\gamma_\mu[
(\PL \URLmu{x}+\PR\ULRmu{x})\psi_{x+\muhat}\nn\\
&&-(\PL \UdaggerRLmu{x-\muhat}+\PR\UdaggerLRmu{x-\muhat})\psi_{x-\muhat}\nn\\
&&-2(\PL \PhiRLmu{x}+\PR\PhiLRmu{x})\psi_{x}]\nn\\
&+&\sum_p[\beta_L\tr U^L(p)+\beta_R\tr U^R(p)]\nn\\
&+&\sum_{x\mu}\tr(1-\PhiLRmu{x}\PhidaggerLRmu{x})\nn
\eea
where $\ULmu{}$ is the usual link variable, $p$ denotes plaquettes, and
\bea
\lefteqn{\ULRmu{x}=W^L_{\mu x}W^{R\dagger}_{-\mu x+\muhat},}\\
\lefteqn{\URLmu{x}=W^R_{\mu x}W^{L\dagger}_{-\mu x+\muhat},}\\
\lefteqn{\PhiRLmu{x}=\half(W^R_{\mu x}W^{L\dagger}_{\mu x}+
W^R_{-\mu x}W^{L\dagger}_{-\mu x})=\PhiRLmu{x},} \nn
\eea
are chiral changing link or site variables.  As a first test, Zenkin has
reproduced the chiral Schwinger model effective action\cite{Jackiw} by
integrating out the extra degrees of freedom in the two dimensional theory.
However, the difficult questions are still to be asked, in light of what we
have learned from other models. For example, what is the spectrum of the
model in four dimensions?  Do bound states form that pair up with the chiral
fermions to result in
a vector-like theory?  This seems likely due to the presence
of the Wilson term which must be considered a strong coupling in the present
context. Further, what is the relation of the gauge invariance in the larger
space of states to gauge invariance in the usual sense.

\subsection{Zaragosa fermions}

Yet another proposal which was discussed at the Rome workshop is
the method proposed by the Zaragosa group\cite{Zaragosa}. The basic idea of
the proposal is to decouple the doubler or replica fermions by suppressing
interactions. The decoupling mechanism is accomplished by replacing all fermion
fields in interaction terms of the action by fields which are
averaged over a lattice hypercube:
\be
\psi^\prime(x) = \frac{1}{16}\sum_b \psi(x+b).
\ee
In momentum space this becomes
\be
\psi^\prime(p) = \prod_\mu \cos \frac{p_\mu}{2} \psi(p),
\ee
which shows us how the decoupling works. In the latter equation, the cosine
is near unity for fermions near the origin in momentum space, but when any
component of momentum gets close to a corner of the Brillouin zone, this
function vanishes, just when a contribution from doublers would arise. So in
principle one can simply write down an action with naive fermions with
standard model quantum numbers, and then replace the $\psi$ fields by
$\psi^\prime$ fields in every interaction term. The desired effect is that the
doublers remain present, but simply as free particles. In
fact, the action is invariant under a generalized fermion shift
symmetry\cite{Zaragosa}
%
%
which guarantees the decoupling of the
$15$ doubler fermions, in the same way that the right-handed neutrino decouples
in models with the usual shift symmetry\cite{Usdecouple}.

The introduction of $\psi^\prime$ into the action breaks chiral symmetry. So
just as in other approaches where this occurs, there is a benefit and a
drawback. The benefit is that the Banks problem is not present, but the
drawback is that as in other models, counterterms need be added to
restore gauge invariance. Results to date include some perturbative
calculations in chiral Yukawa models and some work on the phase diagram, but
so far only limited progress has been made in the direction of including
gauge fields into the theory. We will therefore have to be patient to wait
for further results.

\section{CONCLUSION}

We have learned a lot this year concerning attempts to put chiral gauge
thories on the lattice. Although some proposals must be abandoned, others
have stepped in to take their place and there is is still much to be
understood. So the field looks promising for the year to come.

\section{ACKNOWLEDGMENTS}

I would like to thank many people for discussions, especially, M.\
Golterman, J.\ Smit, G.\ Bodwin, E.\ Kovacs, D.\ Kaplan, S.\ Zenkin and U.\
Heller, and also L.\ Maiani, M.\ Testa, J.\ Vink, S.\ Aoki, I.\ Montvay, K.\
Jansen, H.\ Neuberger, Y.\ Kikukawa, T.\ Banks, P.\ Vranas and W.\ Bock.
Particularly I thank the last two for providing me with figures 1
and 3 respectively to include in these proceedings. I would also like to
thank the organizing committee for a very enjoyable conference and stay in
Amsterdam. This work is funded in part by a grant from the Department of
Energy.


\ninrm
\addtolength{\baselineskip}{-.11\baselineskip}

\newcommand{\NPB}[1]{Nucl.\ Phys.\ {\ninbf B#1}}
\newcommand{\NPBP}[1]{Nucl.\ Phys.\ {\ninbf B} (Proc.\ Suppl.) {\ninbf #1}}
\newcommand{\PRD}[1]{Phys.\ Rev.\ {\ninbf D#1}}
\newcommand{\PLB}[1]{Phys.\ Lett.\ {\ninbf #1B}}
\newcommand{\PRL}[1]{Phys.\ Rev.\ Lett.\ {\ninbf #1}}
\newcommand{\RomeWorkshop}{ presented at the Topical Workshop
on Non-Perturbative Aspects of Chiral Gauge Theories,  Rome, Italy,
March, 1992, to be published in the proceedings}
\newcommand{\Romeibid}{ Rome Workshop, ibid.}


\begin{thebibliography}{999}

\bibitem{Heller}
U.\ M.\ Heller, H.\ Neuberger and P.\ Vranas,
\PLB{283} (1992) 335; preprint FSU-SCRI-92-99;
with M.\ Klomfass, preprint FSU-SCRI-92-150 (hep-lat/9210026);
P.\ Vranas, these proceedings. See also J.\ Kuti, these
proceedings.
\bibitem{Kieu}
T.\ D.\ Kieu, these proceedings.
\bibitem{Montvay}
I.\ Montvay, \PLB{199} (1987) 89; preprint DESY-92-092, June, 1992,
\RomeWorkshop\ and references therein.
\bibitem{Phasediagram}
See the following reviews:
I.\ Montvay, \NPBP{26} (1992) 57; J.\ Shigemitsu, \NPBP{20} (1991) 515;
\MFLG, \NPBP{20} (1991) 528, and refs.\ therein.
\bibitem{DugRand}
M.\ J.\ Dugan and L.\ Randall, \NPB{382} (1992) 419;
preprint MIT-CTP-2115, June 1992, \Romeibid
\bibitem{LueWei}
L\"uscher and Weisz,
\NPB{290} (1987) 25; \NPB{295} (1988) 65; \NPB{318} (1989) 705.
\bibitem{Kuti}
J.\ Kuti, L.\ Lin and Y.\ Shen, \PRL{61} (1988) 678;
in {\ninit Lattice Higgs Workshop} (Berg et.\ al., eds.), World Scientific,
(1988) 140.
\bibitem{Hellerearly}
G.\ Bhanot, K.\ Bitar, U.\ M.\ Heller and H.\ Neuberger,
\NPB{343} (1990) 467; \NPB{353} (1991) 551, erratum {\ninbf B375} (1992) 503.
\bibitem{Banks}
Banks, \PLB{272} (1991) 75;
T.\ Banks and A.\ Dabholkar, preprint RU-92-09, April, 1992.
\bibitem{Karstenearly}
L.\ H.\ Karsten,
in {\ninit ``Field Theoretical Methods
in Particle Physics''} (edited by W.\ R\"uhl), {\sl Plenum Press, New York};
(1980) {\ninit (Kaiserslautern 1979)}.
J.\ Smit, \NPB{175} (1980) 307.
\bibitem{SmitZak}
\JS, Acta Physica Polonica {\ninbf B17} (1986) 531;
\bibitem{Swift}
P.\ D.\ V.\ Swift, \PLB{145} (1984) 256.
\bibitem{Usdecouple}
\MFLG\ and \DNP, \PLB{225} (1989) 159.
\bibitem{BockpointA}
W.\ Bock, A.\ K.\ De, C.\ Frick, J.\ J\'ersak and T.\ Trappenberg,
\NPB{378} (1992) 652.
\bibitem{MaianiRome}
L.\ Maiani, \Romeibid
\bibitem{UsRome}
\MFLG\ and \DNP, preprint Wash.\ U.\ HEP/92-81 (hep-lat/9207004),
\Romeibid
\bibitem{FunKasAoki}
S.\ Aoki, \PRD{38} (1988) 618;
K.\ Funakubo and T.\ Kashiwa, \PRL{60} (1988) 2113; \PRD{38} (1988) 2602.
\bibitem{SmitSeillac}
\JS, \NPBP{4} (1988) 451.
\bibitem{JanandUs}
\MFLG, \DNP\ and \JS, \NPB{370} (1992) 51.
\bibitem{BockJapan}
W.\ Bock, \NPBP{26} (1992) 220.
\bibitem{Bockcharged}
W.\ Bock, A.\ K.\ De and \JS, preprint HLRZ-91-81, J\"ulich; W.\ Bock,
\NPBP{26} (1992) 220.
\bibitem{oneoverw}
\MFLG\ and
\DNP, \NPBP{20} (1991) 577; \MFLG,
\DNP\ and E.\ Rivas, \NPB{377} (1992) 405.
\bibitem{Aokilat92}
S.\ Aoki, these proceedings.
\bibitem{Aokicouplings}
S.\ Aoki, preprint UTHEP-241, July, 1992.
\bibitem{UsJapan}
\MFLG\ and \DNP, \NPBP{26} (1992) 483.
\bibitem{Bock2d}
W.\ Bock, A.\ K.\ De, E.\ Focht and \JS, preprint ITFA-92-21 (hep-lat/9210022).
\bibitem{DuganMan}
M.\ J.\ Dugan and A.\ V.\ Manohar, \PLB{265} (1991) 137.
\bibitem{EichPres}
E.\ Eichten and J.\ Preskill, \NPB{268} (1986) 179.
\bibitem{Kutietal}
A.\ Hasenfratz, P.\ Hasenfratz, K.\ Jansen, J.\ Kuti and Y.\ Shen,
\NPB{365} (1991) 79.
\bibitem{UsEichPres}
\MFLG, \DNP\ and E.\ Rivas, preprint Wash.\ U.\ HEP/91-61,
to be published in \NPB{};
preprint Wash.\ U.\ HEP/92-82 (hep-lat/9207005), \Romeibid
\bibitem{Roma}
A.\ Borrelli, L.\ Maiani, G.C.\ Rossi, R.\ Sisto and M.\ Testa,
\PLB{221} (1989) 360; \NPB{333} (1990) 355.
\bibitem{Rometwoloop}
G.\ C.\ Rossi, R.\ Sarno and R.\ Sisto, preprint ROME-888-1992, May, 1992.
\bibitem{SmitRome}
\JS, \NPBP{26} (1992) 480;
\JS, \Romeibid
\bibitem{BodKov}
G.\ T.\ Bodwin and E.\ V.\ Kovacs, preprint ANL-HEP-CP-92-103
(hep-lat/9211021),
these proceedings.
\bibitem{Gockeler}
M.\ Gockeler and G.\ Schierholz, preprint HLRZ-92-33, June, 1992,
\Romeibid; M.\ Gockeler, these proceedings.
\bibitem{AlvDP}
Alvarez-Gaum\'e, S.\ Della Pietra and V.\ Della Pietra, \PLB{166B}
(1986) 177.
\bibitem{Kikukawa}
Y.\ Kikukawa, preprint DPNU-91-48, November 1991.
\bibitem{RomeMaj}
L.\ Maiani, G.\ C.\ Rossi and M.\ Testa, \PLB{292} (1992) 397.
\bibitem{Pryor}
C.\ Pryor, \PRD{43} (1991) 2669.
\bibitem{Staggered}
T.\ Banks, S.\ Raby, L.\ Susskind, J.\ Kogut, D.\ R.\ T.\ Jones, P.\ N.\
Scharbach
and D.\ Sinclair, \PRD{15} (1976) 1111; L.\ Susskind,
\PRD{16} (1977) 3031.
\bibitem{KawaSmit}
N.\ Kawamoto and \JS, \NPB{192} (1981) 100.
\bibitem{STWeisz}
H.\ S.\ Scharatchandra, H.\ J.\ Thun and P.\ Weisz, \NPB{192} (1981) 205.
\bibitem{StaggeredMC}
W.\ Bock, J.\ Smit and J.\ C.\ Vink, \PLB{291} (1992) 297;
W.\ Bock, J.\ Smit C.\ Frick and J.\ C.\ Vink, preprint ITFA 92-23,
August, 1992.
\bibitem{definestaggered}
C.\ van\ den\ Doel and \JS, \NPB{228} (1983) 122;
\MFLG\  and \JS, \NPB{245}(1984) 61;  \NPB{255}(1985) 328;
\MFLG, \NPB{273} (1986) 663; \NPB{278} (1986) 417;
H.\ Joos and M.\ Shaefer, Z.\ Phys.\ {\ninbf C34} (1987) 465.
\bibitem{Bocklat92}
W.\ Bock, these proceedings; J.\ Vink, these proceedings.
\bibitem{Montvayetal}
C.\ Frick, L.\ Lin, I.\ Montvay, C.\ M\"unster, M.\ Plagge, T.\ Trappenberg and
H.\ Wittig, preprint hep-lat/9207021.
\bibitem{CalHarv}
C.\ G.\ Callan and J.\ A.\ Harvey,
\NPB{250} (1985) 427.
\bibitem{Kaplan}
D.\ B.\ Kaplan, \PLB{288} (1992) 342.
\bibitem{SchmaltzJansen}
K.\ Jansen  and M.\ Schmaltz, preprint  UCSD-PTH-92-29 (hep-lat/9209002).
\bibitem{GoldWil}
J.\ Goldstone and F.\ Wilczek, \PRL{47} (1981) 986.
\bibitem{GolJanKap}
\MFLG, K.\ Jansen and D.\ B.\ Kaplan,
UCSD-PTH-92-28, Aug., 1992.
\bibitem{Jansen2d}
K.\ Jansen \PLB{288} (1992) 348.
\bibitem{BodKovPriv}
I thank G.\ Bodwin and E.\ Kovacs for pointing this out to me.
\bibitem{Zenkin}
S.\ V.\ Zenkin, preprint DFTUZ-92-7 (Moscow, INR) (hep-lat/9205012), May, 1992.
\bibitem{Brydges}
D.\ Brydges, J.\ Fr\"ohlich and E.\ Seiler, Ann.\ Phys.\ 121 (1979) 227.
\bibitem{Jackiw}
R.\ Jackiw and R.\ Rajaraman, \PRL{54} (1985) 1219.
\bibitem{Zaragosa}
J.\ L.\ Alonso, J.\ L.\ Cortes, P.\ Boucaud and E.\ Rivas, \PLB{201} (1988)
340; \PRD{40} (1989) 4123; \PLB{237} (1990) 476; \PRD{44} (1991) 3258 ;
with F.\ Lesmes, preprint DFTUZ-92-8 (hep-lat/9206005), \Romeibid

\end{thebibliography}
\end{document}